\documentclass[prl,aps,twocolumn,showpacs,floatfix,nofootinbib,superscriptaddress]{revtex4}
\usepackage[pdftex]{graphicx}% Include figure files
\graphicspath{{figures/pdf/}}
\DeclareGraphicsExtensions{.pdf}
\usepackage{dcolumn}% Align table columns on decimal point
\usepackage{bm}% bold math
\usepackage{amsthm,amsmath,amssymb}
\def\vec#1{{\bm #1}}
\def\op#1{#1}

\def\ket#1{| #1 \rangle}
\def\bra#1{\langle #1 |}

\def\norm#1{\| #1 \|}

\def\Tr{\operatorname{Tr}}

\def\D{\mathcal{D}}

\def\L{\mathcal{L}}

\def\up{{\uparrow}}
\def\down{{\downarrow}}
\def\s#1{\op{\sigma}_{#1}}
\def\oo{\omega}
\def\eps{\epsilon}

\def\ss{\rm ss}

\begin{document}
\title{Generating maximal entanglement between non-interacting atoms \\
by collective decay and symmetry breaking}

%\title{Arbitrarily high concurrence, pure steady-state entanglement
%between non-interacting atoms via collective decay and adiabatic
%control}

\author{Xiaoting Wang}
\affiliation{Department of Applied Maths and Theoretical Physics,
University of Cambridge, Wilberforce Road, Cambridge, CB3 0WA,
United Kingdom}

\author{S. G. Schirmer}
\affiliation{Department of Applied Maths and Theoretical Physics,
University of Cambridge, Wilberforce Road, Cambridge, CB3 0WA, United
Kingdom}

%% based on Jan 17 version
\date{\today}

\begin{abstract}
A simple scheme is presented for achieving effectively maximal
pure-state entanglement between non-interacting atoms through purely
collective decay and controlled symmetry breaking.  The scheme requires
no measurements or feedback or even knowledge of the initial states of
the atoms.  It relies on breaking the symmetry of the system Hamiltonian
to ensure the existence of a unique attractive steady state and minimal
control to achieve almost perfect overlap of this steady state with the
maximally entangled singlet state.  We demonstrate how our scheme can be
implemented for two qubits encoded in hyperfine levels of atoms such as
Rubidium in a lossy microwave cavity using only small magnetic field
gradient.  Error analysis suggests considerable robustness with regard
to many imperfections including atomic decay, asymmetric atom-cavity
coupling and frequency offsets.
\end{abstract}

\maketitle

Entanglement is an important resource in quantum physics, especially
quantum information, and many schemes to generate entanglement have been
proposed, using various tools including coherent control, measurements
and feedback.  In coherent control schemes
(e.g.,~\cite{coherent-control-entanglement}) dissipation tends to
destroy entanglement created by Hamiltonian evolution and reduce the
control effectiveness, but it can also be a resource for creating
entanglement.  Some interacting systems have entangled ground states and
relaxation processes can drive the system into such states, known as
cooling into entangled states~\cite{cooling}.  More generally, careful
design of the Hamiltonian and interaction with the environment, i.e.,
Hamiltonian and reservoir engineering, can ensure decay to a certain
state~\cite{eng-open-dynamics,steady-state-eng,qph0909_1596}.  Reservoir
engineering by measurements has been used to drive systems into
entangled states~\cite{measurement-induced-entanglement}, and schemes
relying on direct or indirect feedback have been proposed
\cite{feedback-Wiseman}.  Entanglement induced purely by interaction
with an environment has also been
explored~\cite{environment-induced-entanglement}, but the entanglement
achieved this way is usually low-grade mixed-state entanglement and
often transient, i.e., the system relaxes to a separable steady state
eventually.

Ideally, we would like to engineer systems that relax to a stable,
maximally entangled, pure steady state regardless of the initial state.
For open systems described by Lindblad master equation this can be
achieved by engineering the dynamics to render the steady state unique
as uniqueness implies that any initial state ``collapses'' to this
state~\cite{steady-state-eng,qph0909_1596}.  Based on this idea, we show
that effectively maximal pure-state entanglement can be \emph{created}
for atoms in a damped cavity driven by a classical field, even if there
is \emph{no interaction} between the atoms and no measurements are
performed, i.e., \emph{purely by dissipation}, and the initial state is
unknown.  Of course, the singlet state for two identical atoms in a
cavity is maximally entangled, but it is not attractive, i.e., if the
system is initially in any other state, it will \emph{not} evolve into
the latter.  However, breaking the symmetry in the model, e.g., by
varying the energy level splitting of the atoms slightly, results in a
unique attractive steady state, and surprisingly, this is a pure state
whose concurrence can be made arbitrarily close to $1$ by suitable
choice of the parameters, and which can be reached in a short time from
an unknown initial state simply by adiabatically decreasing the induced
asymmetry.

\begin{figure}
\includegraphics[width=\columnwidth]{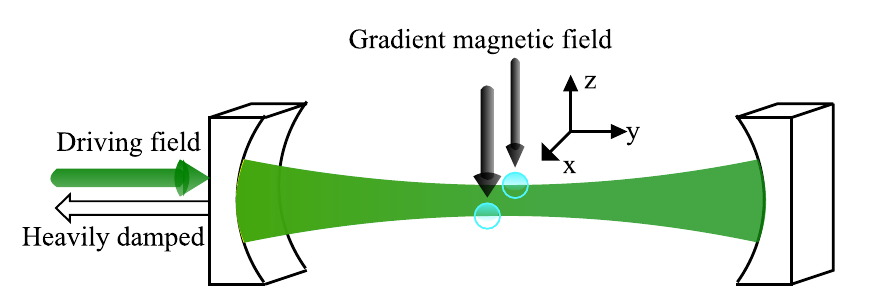} \caption{Two trapped
atoms in a heavily damped cavity.  The cavity field is slaved to the
driving field.  The quantization axis and field polarization direction
are defined by a static magnetic field along $z$-axis.  A small magnetic
field gradient along $x$-axis permits variation of the atomic energy
levels.} \label{fig:setup}
\end{figure}

There are many possible physical realizations for the type of system
described but to fix the ideas and demonstrate the feasibility of the
scheme we consider two trapped ultracold atoms such as $^{87}$Rb,
situated in a damped microwave cavity driven by a classical field as
shown in Fig.~\ref{fig:setup} as a particularly promising
implementation.  A magnetic field is applied to provide a quantization 
axis and photon polarization direction, and the qubit states are encoded 
in two hyperfine magnetic sublevels such as
$\ket{\up} \equiv\ket{F=1,m_F=1}$ and
$\ket{\down}\equiv\ket{F=2,m_F=2}$.  These basis states are stable with
very long coherence times, i.e., once the atoms are prepared in a state
such as the singlet state, we can expect them to remain in this state
for a long time.  The system could be prepared by loading the ultracold
atoms into optical lattices to ensure the atoms are well localized and
stationary.  A detuning of the atomic transition frequencies of
$\pm\Delta\oo$ from the cavity field frequency $\oo$ can be achieved via
the Zeeman effect by applying a small magnetic field gradient.  The
total Hamiltonian of the system is
$H_{tot}=H_{a}+H_{c}+H_{aa}+H_{ac}+H_{dc}$, where the Hamiltonians for
the two atoms and the cavity are
\( H_{a}=\hbar (\oo_1\s1^\dag\s1 + \oo_2\s2^\dag\s2), \)
\( H_{c} = \hbar\omega b^\dagger b \)
and $\sigma_n$ $(n=1,2)$ and $b$ are the annihilation operators for the
$n$th atom and intra-cavity field, respectively.  In a typical setup
with neutral atoms at least several microns apart direct interactions
are negligible, $H_{aa}=0$.  The atom-cavity interaction is given by
\( H_{ac} = \hbar[g_1(\s1+\s1^\dag)+g_2(\s2+\s2^\dag)] \otimes (b+b^\dag) \)
and the classical coherent driving field $f(t)=A\cos(\omega t)$ adds the
term
\(  H_{dc}=i A d \cos(\omega t) (b-b^\dag), \)
where $d$ is the coupling strength.  Including spontaneous emission and
decay of the cavity field, the evolution of the total density operator
$\rho_t$ of the driven atom-cavity system is given by 
\begin{equation}
 \label{eq:LB1}
  \dot{\rho}_t = -\tfrac{i}{\hbar}[H_{tot},\rho_t]
      +\gamma_1 \D[\s1]\rho_t + \gamma_2\D[\s2]\rho_t
      +\gamma_b \D[b]\rho_t,
\end{equation}
with $\D[a]\rho=a\rho a^\dag -\tfrac{1}{2}(a^\dag a\rho + \rho
a^\dag a)$, which can be simplified by transforming to a rotating
frame (RFT) $\rho_t'=U_0(t)^\dag \rho_t U_0(t)$ with %%
\( U_0(t)=e^{-i\oo t(\s1^\dag\s1+\s2^\dag\s2 +b^\dag b)}.  \)
The RFT leaves the Lindblad terms invariant and gives
\begin{equation}
 \label{eq:LB2}
  \dot{\rho}_t' = -\tfrac{i}{\hbar}[H_{tot}',\rho_t']
      +\gamma_1\D[\s1]\rho_t' +\gamma_2\D[\s2]\rho_t'
      +\gamma_b\D[b]\rho_t'.
\end{equation}
In the small detuning limit, $|\omega_n-\omega|\ll 2\omega$, we can
neglect terms rotating with frequency $2\omega$, obtaining
\begin{equation} %\textstyle
  \tfrac{1}{\hbar}H_{tot}'
  = i\alpha_0 (b-b^\dag) + H_z(\Delta\oo_n) + g(J b^\dag+J^\dag b),
\end{equation}
where $\alpha_0=Ad/2\hbar$ is the Rabi frequency of the driving field,
$J=g^{-1}(g_1 \s1+g_2\s2)$ the collective atom-cavity interaction, and
$H_z(\Delta\omega_n)=\Delta\omega_1\sigma_1^\dag\sigma_1+\Delta\omega_2
\sigma_2^\dag\sigma_2$ with $\Delta\omega_n=\omega_n-\omega$
incorporates the detuning of the atoms from the cavity field.  The RFT
and $J$ above differ slightly from the usual interaction picture
transformation and definition of $J$ to allow for off-resonant and
asymmetric interactions.
We further eliminate $i\alpha_0(b-b^\dag)$ by the standard displacement
transformation $\rho_t''= U_b(\beta)\rho_t'U_b(\beta)^\dag$ with
$U_b(\beta)=e^{\beta(b-b^\dag)}$.  $U_b(\beta)$ commutes with the atomic
operators $\s{n}$ and the standard commutation relations
$[b-b^\dag,b]=[b,b^\dag]=1$ give $U_b(\beta) a U_b(\beta)^\dag=a+\beta$
for $a=b$ or $b^\dag$ (e.g., using Hadamard's lemma), and thus 
\begin{align*}
 H_{tot}''       &= U_b(\beta)H_{tot}'U_b(\beta)^\dag = H_{tot}'+\hbar g\beta J_x,\\
 \D[b'']\rho_t'' &= \D[b+\beta]\rho_t'' =
            \D[b]\rho_t''+\tfrac{\beta}{2}[b-b^\dag,\rho_t''],
\end{align*}
with $J_z=J+J^\dag$.  Setting $\beta=-2\tfrac{\alpha_0}{\gamma_b}$
cancels  driving field term and gives the simplified master equation
\begin{equation}
 \label{eq:LB3}
 \dot{\rho}_t''
  = \L(\rho_t'')-i\lambda\gamma_b [J b^\dag+J^\dag b,\rho_t'']
    +\gamma_b \D[b]\rho_t'',
\end{equation}
where the Lindblad operator for the atomic subsystem is
\begin{equation}
  \label{eq:La} \textstyle
  \L(\rho_t'') = -i[H_z(\Delta\omega_n) + \alpha J_x,\rho_t'']
                 + \sum_n \gamma_n \D[\sigma_n]\rho_t''
\end{equation}
and $\lambda=g/\gamma_b$, $\alpha=\beta g = -2\alpha_0\lambda$.  If the
cavity is heavily damped following the standard procedure for adiabatic
elimination of the cavity mode (e.g.~\cite{feedback-Wiseman,tanas})
leads to the equation of motion for the atomic subsystem
\begin{equation}
\label{eq:LB4}
 \dot{\rho}_t^a \approx
 \L(\rho_t^a) + \gamma \D[J]\rho_t^a +(\D[J^\dag]-\D[J])
 \bra{1}\rho_t\ket{1}
\end{equation}
with $\gamma=4\lambda^2\gamma_b=4g^2/\gamma_b$.  If the population of
the first excited cavity mode $\ket{1}$, and thus
$\bra{1}\rho_t\ket{1}$, remains small and the atomic spontaneous
emission rates $\gamma_n\ll\gamma$ then the contributions of
$(\D[J^\dag]-\D[J])\bra{1}\rho_t\ket{1}$ and
$\gamma_n\D[\sigma_n]\rho_t^a$ will be negligible, yielding the
Dicke-type master equation for the atomic subsystem with a detuning term
\begin{equation}
 \label{eq:LB5}
  \dot{\rho}_t^a =-i[H_z(\Delta\oo_n)+\alpha J_x,\rho_t^a]+\gamma\D[J]\rho_t^a.
\end{equation}
If the detuning of the atoms from the cavity frequency $\omega$ and
their interaction with the cavity field are symmetric, i.e.,
$\Delta\omega_1=\Delta\omega=-\Delta\omega_2$ and $g_1=g_2=g$, then the
detuning term $H_z(\Delta\omega_n)\equiv \Delta\omega J_z$, and it is
easy to verify that
\( \ket{\Psi_{\ss}}=
  \Omega^{-1}(\Delta\omega \ket{\up\up} + \alpha\ket{\up\down} -\alpha \ket{\down\up})
\)
with $\Omega=\sqrt{\Delta\omega^2+2\alpha^2}$ is an eigenstate with
eigenvalue $0$ of the effective Hamiltonian $H=\Delta\omega J_z+\alpha
J_x$ and the Lindblad operator $J$ and thus a steady state of the
dynamics (\ref{eq:LB5}).  For $\Delta\omega=0$ (no detuning), the
populations of the $0$ and $1$ eigenspaces of $J_x$ are conserved, i.e.,
the system is decomposable and there are infinitely many other steady
states, rendering $\ket{\psi_{\ss}}$ non-attractive~\cite{qph0909_1596},
i.e., if the atoms are initially in a state other than the maximally
entangled singlet state (MES) $\ket{\psi_{\ss}}$, they will \emph{not}
evolve into the MES but a low-concurrence mixed state~\cite{Schneider},
i.e., the MES is protected against the decay but cannot be created
starting from another state this way.

\begin{figure}
\includegraphics[width=\columnwidth]{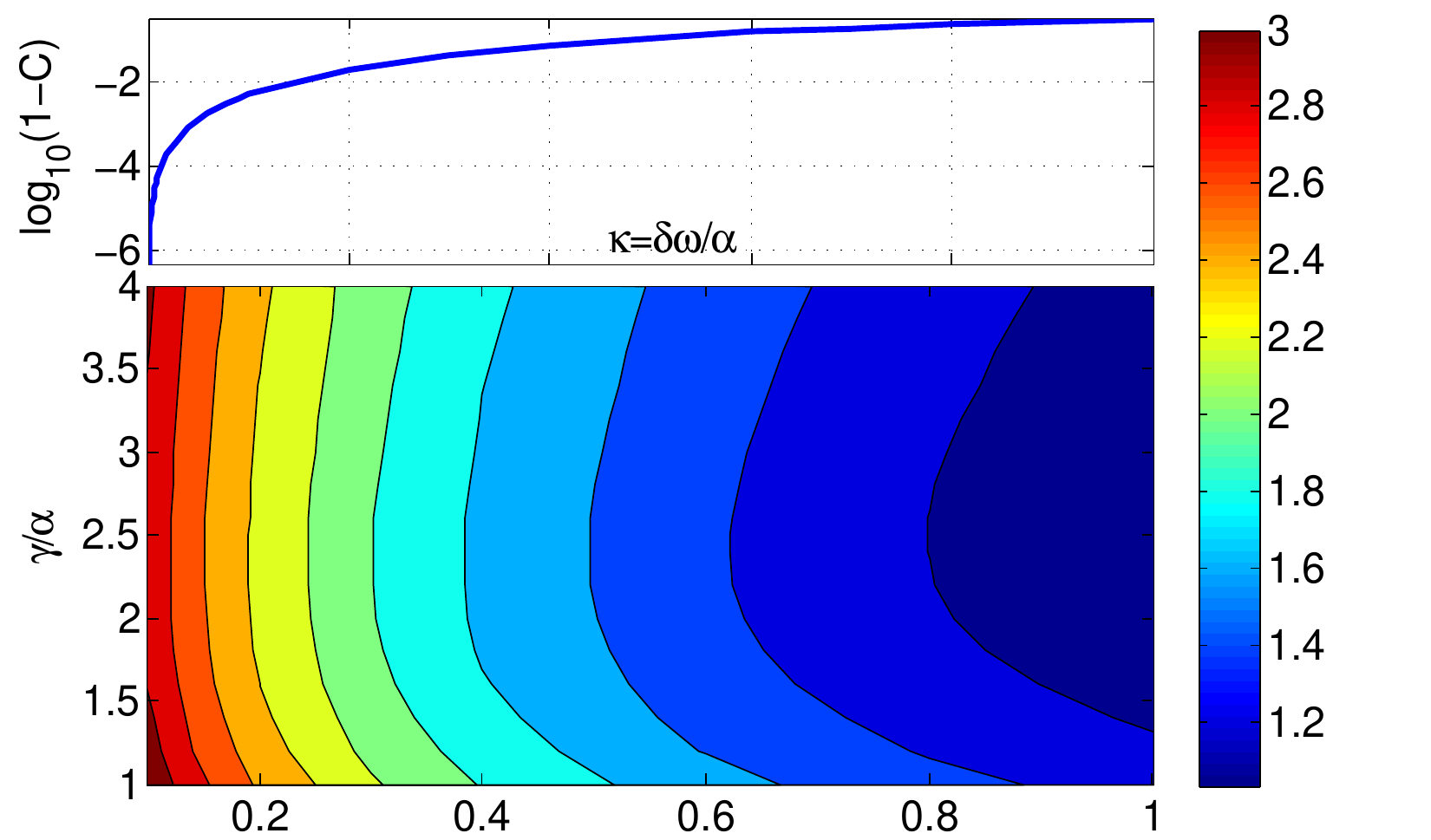} %%%%
\caption{(Color online): Steady state concurrence $C$ as function of
$\kappa$ (top) and $\log_{10}$ of time needed to reach 99\% of the
steady state concurrence (bottom) as function of $\gamma/\alpha$ and
$\kappa$.}  \label{fig2}
\end{figure}

To be able to create a MES, or a state arbitrarily close to it, we must
ensure that the target state is the only steady state of the
system. This can be accomplished by noting that even a small detuning
breaks the symmetry and renders $\ket{\psi_{\ss}}$ the \emph{unique
globally attractive} steady state~\cite{qph0909_1596}.  Moreover, as the
steady state is pure, any stochastic quantum trajectory of the system
must converge to this state.  Setting $\kappa=\Delta\omega/\alpha$, the
concurrence of the steady sate is
$C_{\ss}=C(\Psi_{\ss})=2\alpha^2/\Omega^2=(1+\kappa^2/2)^{-1}$, showing
that attains $1$ only for $\kappa=0$, but can be made arbitrarily close
to $1$ for $\Delta\omega\neq 0$ by increasing
$\alpha$. Fig.~\ref{fig2}(a) shows the dependence of $C_{\ss}$ on the
detuning parameter $\kappa$; 99\% concurrence is achievable with
$\kappa\approx 0.14$.  The steady state concurrence $C_{\ss}$ is
independent of the decay rate $\gamma$ but $\gamma$ determines the rate
of convergence to the steady state.  Fig.~\ref{fig2}(b) shows the
average time required for the concurrence of the system state to reach
99\% of $C_{\ss}$. This time is virtually independent of the initial
state, i.e., the variation for different initial states is minimal, but
depends strongly on $\kappa$, decreasing substantially with increasing
$\kappa$.  The graphs also suggest that there is an optimum value of
$\gamma/\alpha$ between $1.5$ and $2.5$, depending on $\kappa$, with
$\gamma=2\alpha$ being close to optimal for a wide range of $\kappa$.
This suggests that the best strategy to achieve high concurrence in the
shortest possible time is to choose $\gamma/\alpha$ near the optimum
value and the largest possible $\kappa$ that still allows us to reach
the desired concurrence, or better yet, start with a large $\kappa_0$
and gradually decrease it to $0$, e.g., linearly or
exponentially. %$\kappa(t)=\kappa_0 e^{-\mu t}$.

\begin{figure}
\includegraphics[width=\columnwidth]{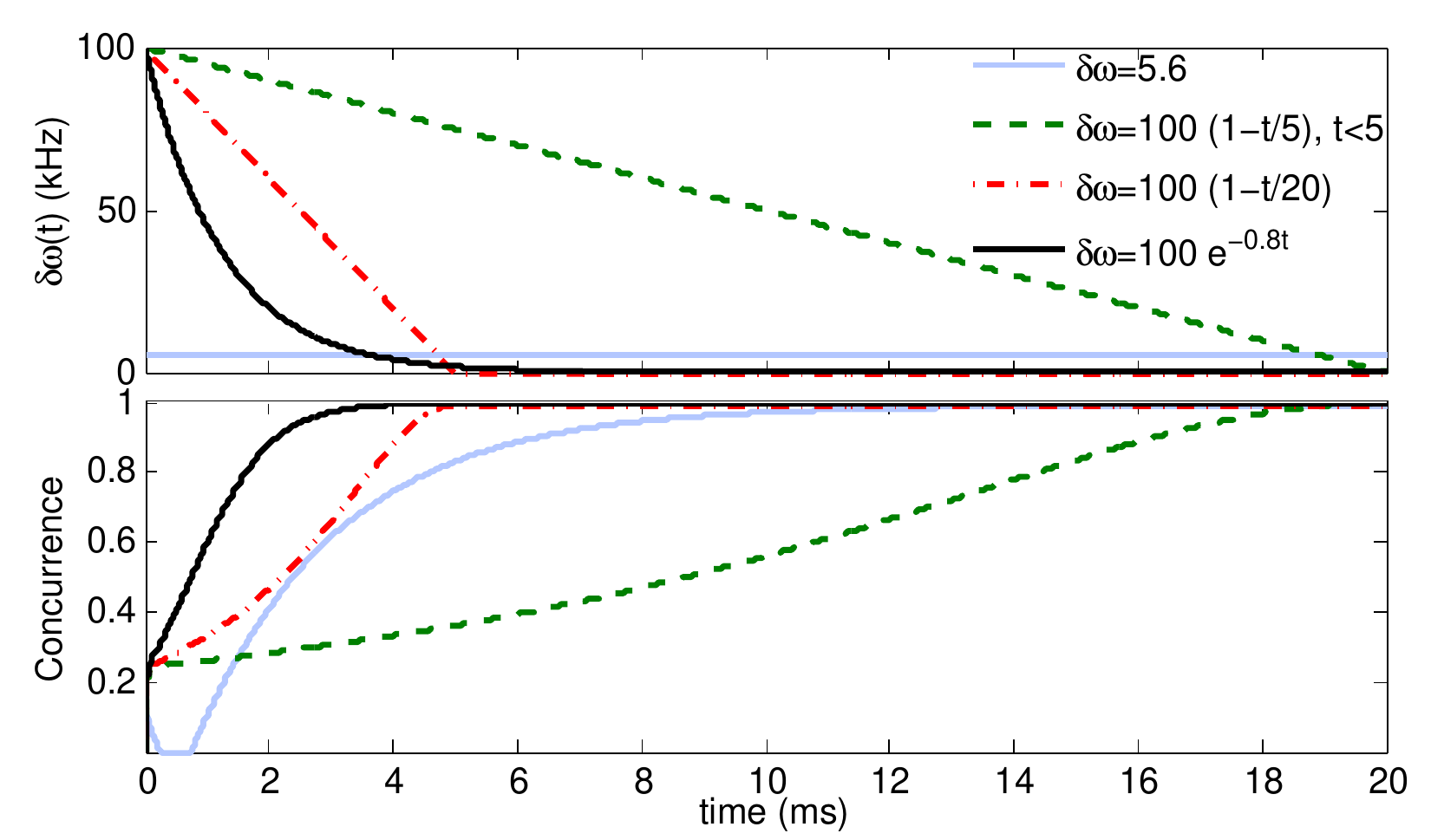}%%%
\caption{(Color online): Concurrence $C(t)$ (bottom) for various
(simple) detuning profiles $\Delta\omega(t)$ (top) shows that
exponential detuning achieves the highest concurrences fastest.}\label{fig3}
\end{figure}

To assess whether we can achieve high entanglement for the setup in
FIG.~\ref{fig:setup} with this scheme, we turn to simulations. For
$^{87}$Rb we choose $\omega=6.83$~GHz, and for atoms a few microns
apart, detunings up to $1$~MHz are attainable via a magnetic field
gradient.  Recent experimental results~\cite{experiment} suggest
that realistic values for the coupling constant $g$ and effective
Rabi frequency for the driving field $\alpha_0$ in a lossy microwave
cavity are on the order of a few hundred kHz, and the cavity damping
rate $\gamma_b$ can easily be made about one order of magnitude
larger.  We choose $\alpha_0=g=200$~kHz and $\gamma_b=10g$, which
gives $\lambda=0.1$, $\gamma=80$~kHz and $\alpha=40$~kHz, and
$\gamma/\alpha=2$, which is in the optimal range to minimize the
convergence time $t_{99}$ according to Fig.~\ref{fig2}(b). Our
previous results show that to achieve a steady state concurrence of
$99$\% requires $\kappa\le 0.14$ or $\Delta\omega\le 5.6$~kHz.
Fig.~\ref{fig3} shows that with a constant detuning
$\Delta\omega=5.6$~kHz we can reach 99\% concurrence from an
arbitrary initial state in less than 20~ms.  The figure also shows
that this result can be significantly improved if we start with a
larger detuning and gradually decrease it to $0$. E.g., starting
with $\Delta\omega=2.5\alpha=100$~kHz, which is still well below the
limit of $1$~MHz, we can obtain over 99\% concurrence in less than
$5$~ms if the detuning is decreased exponentially.

Next we verify the validity of the approximations made in deriving
(\ref{eq:LB5}), i.e., adiabatic elimination and negligibility of the
terms $\gamma_n\D[\sigma_n]\rho_t^a$ and
$(\D[J^\dag]-\D[J])\bra{1}\rho_t\ket{1}$.  Comparison of the exact
solution $\rho_t$ given by (\ref{eq:LB3}) and the solution
$\rho_t^a$ of the approximate model~(\ref{eq:LB5}) suggests that the
latter is a very good approximation in the parameter regime above.
For $\gamma_1=\gamma_2=0$, the reduced density operator for the
atomic subsystem obtained by solving (\ref{eq:LB3}) and tracing over
the cavity modes, $\rho_t^{e}= \Tr_b[\rho_t]$, and the approximate
solution $\rho_t^a$ of (\ref{eq:LB5}) are in excellent agreement
with the norm error $\eps=\max_{t} \norm{\rho_t^e-\rho_t^a} \le
0.0201$ for $\Delta\omega=5.6$~kHz (const) and $\le 0.0172$ for
$\Delta\omega(t)=100 e^{-0.8t}$, and we have verified that the
concurrences predicted by the exact and approximate models are in
very good agreement for all the simulations below.

\begin{figure}
\includegraphics[width=0.48\columnwidth]{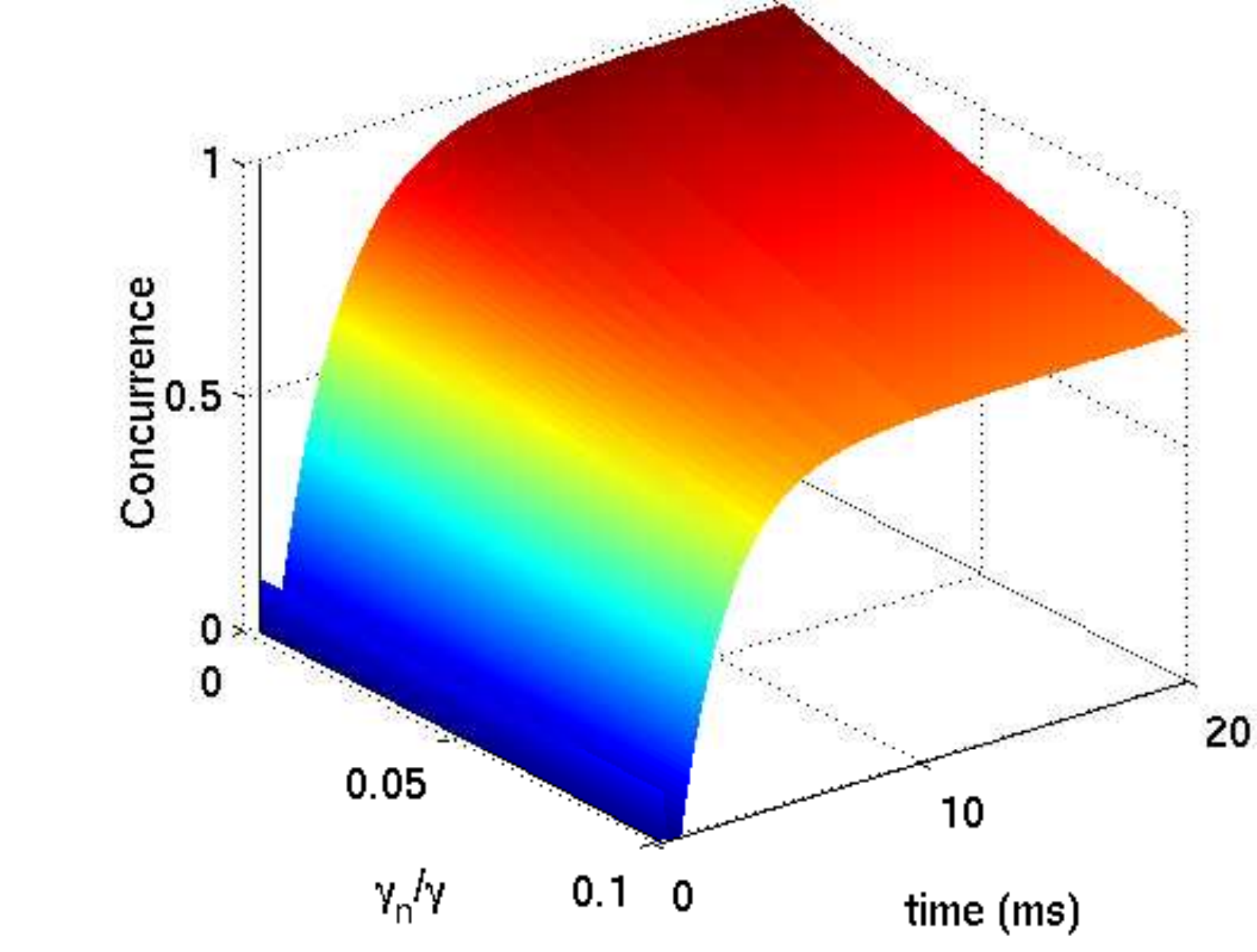}
\includegraphics[width=0.48\columnwidth]{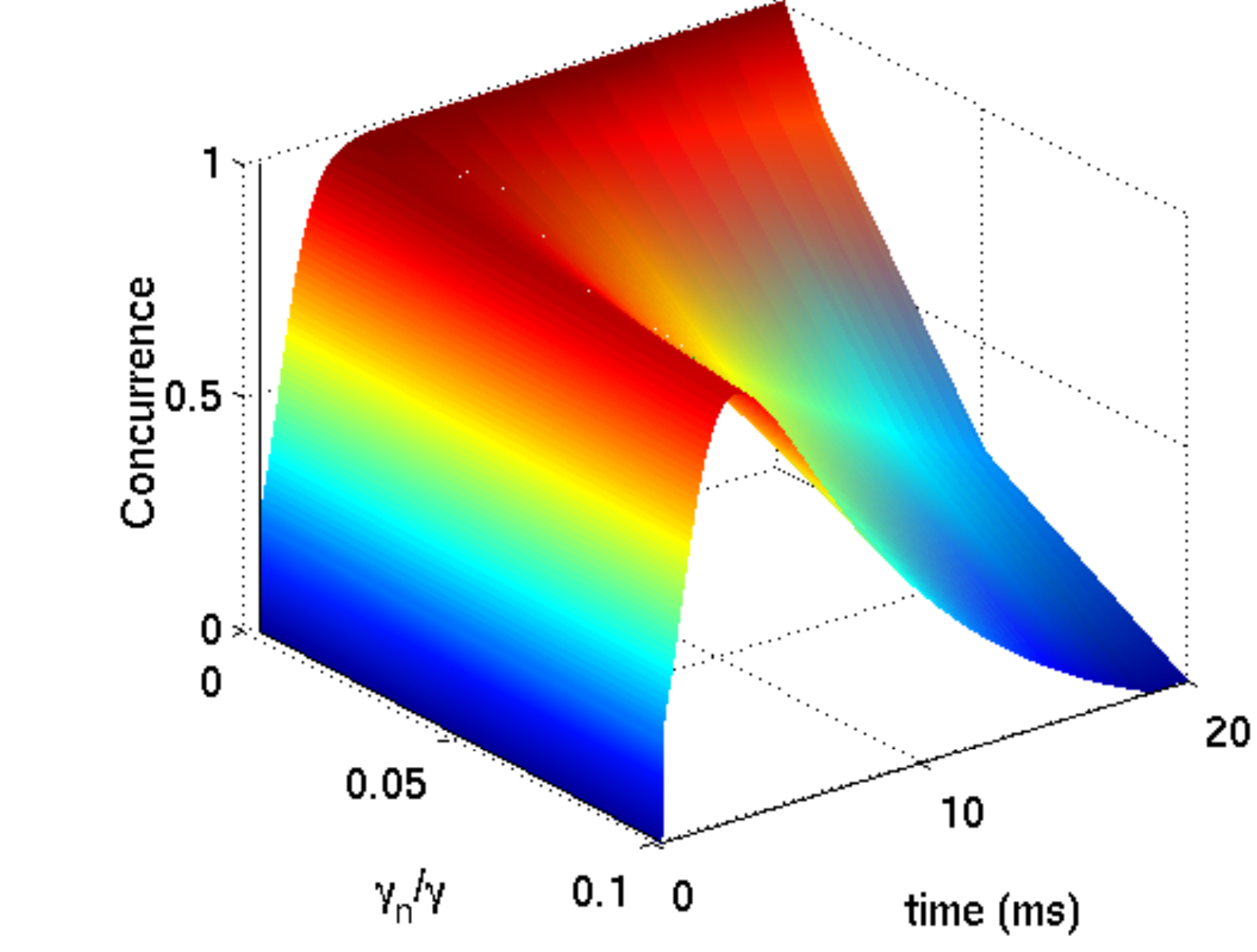}
\caption{(Color online): Concurrence $C(t)$ as a function of time and
atomic decay rates $\gamma_1=\gamma_2$ for $\Delta\omega=5.6$~kHz
constant (left) and exponential detuning $\Delta\omega(t)=100 e^{-0.8t}$
(right).}  \label{fig4}
\end{figure}

\begin{figure}
\includegraphics[width=0.48\columnwidth]{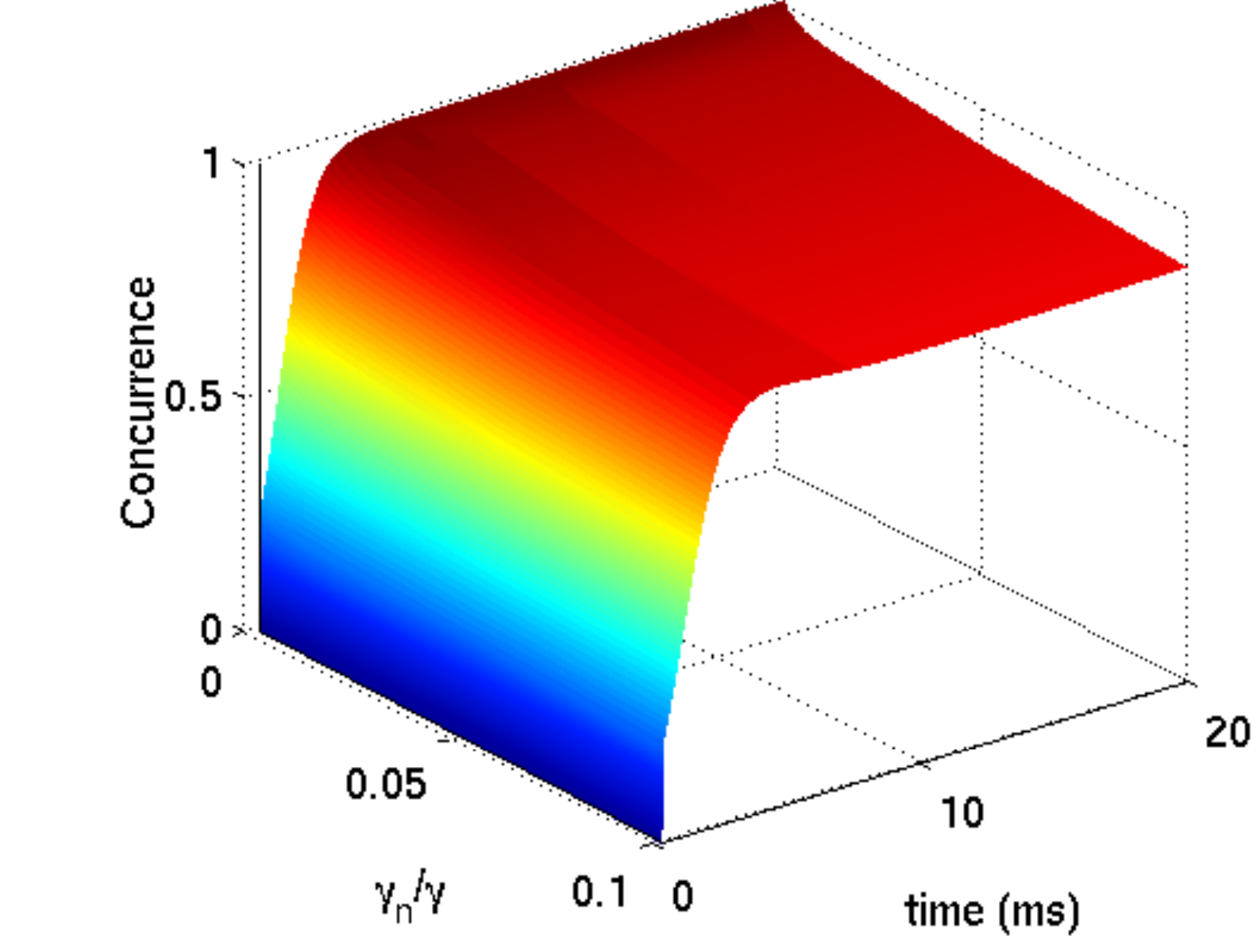}
\includegraphics[width=0.48\columnwidth]{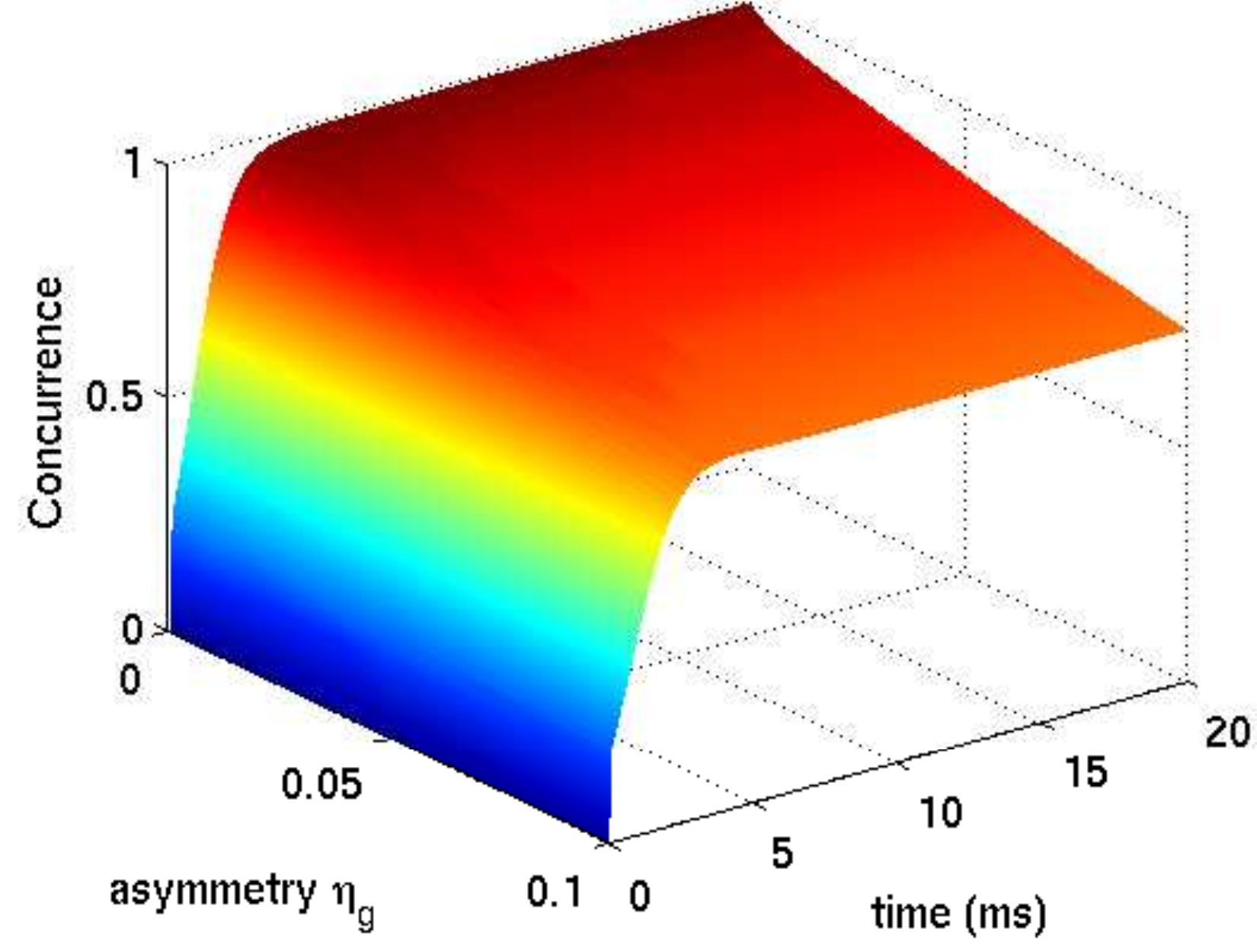}
\caption{(Color online): Concurrence $C(t)$ as a function of time
and the atomic decay rates $\gamma_1=\gamma_2$ (left) and time and
asymmetry in the atom-cavity couplings $\eta_g$ (right) for
exponential detuning with constant offset.} \label{fig5}
\end{figure}

For the systems considered here $\gamma_n\ll \gamma$ should be easily
achievable under current experimental conditions, as the lifetimes of
hyperfine states of ultracold atoms such as $^{87}$Rb are on the order
of seconds, suggesting $\gamma_n<1$~Hz, while our effective cavity decay
rate is $\gamma=80$~kHz, but the effect of non-zero atomic decay rates
on the attainable concurrence deserves investigation.  Fig.~\ref{fig4}
shows the concurrence of the system state as function of time and the
atomic decay rates $\gamma_1=\gamma_2$ for constant and exponentially
decreasing detuning.  For $\gamma_n=0.001$~kHz, the effect of the atomic
decay is virtually imperceptible on the timescales considered but for
$\gamma=0.1$~kHz, atomic decay decreases the attainable concurrence
substantially.  The exponentially decaying $\Delta\omega$ achieves a
higher peak concurrence in the presence of atomic decay, but as the
detuning approaches $0$, atomic decay takes over and drives the
concurrence down again, while for constant detuning the concurrence
reaches a lower steady state value but remains near constant on the
timescales considered.  This might be explained in terms of inhibition
of spontaneous emission by detuning~\cite{Kleppner} resulting in reduced
effective decay rates.  This suggests that in the presence of
non-negligible atomic decay the best strategy is to start with a large
initial detuning to maximize the rate of convergence, but to decrease it
not to zero but a non-zero asymptotic value $\Delta\omega_f$
(offset), chosen such that $2/(2+(\tfrac{\Delta\omega_f}{\alpha})^2)$ is
close to the peak value for the concurrence.  Indeed,
Fig.~\ref{fig5}(left) shows that this strategy appears highly effective
in maximizing the steady-state entanglement and the rate at which it is
reached.

\begin{figure}
\includegraphics[width=0.48\columnwidth]{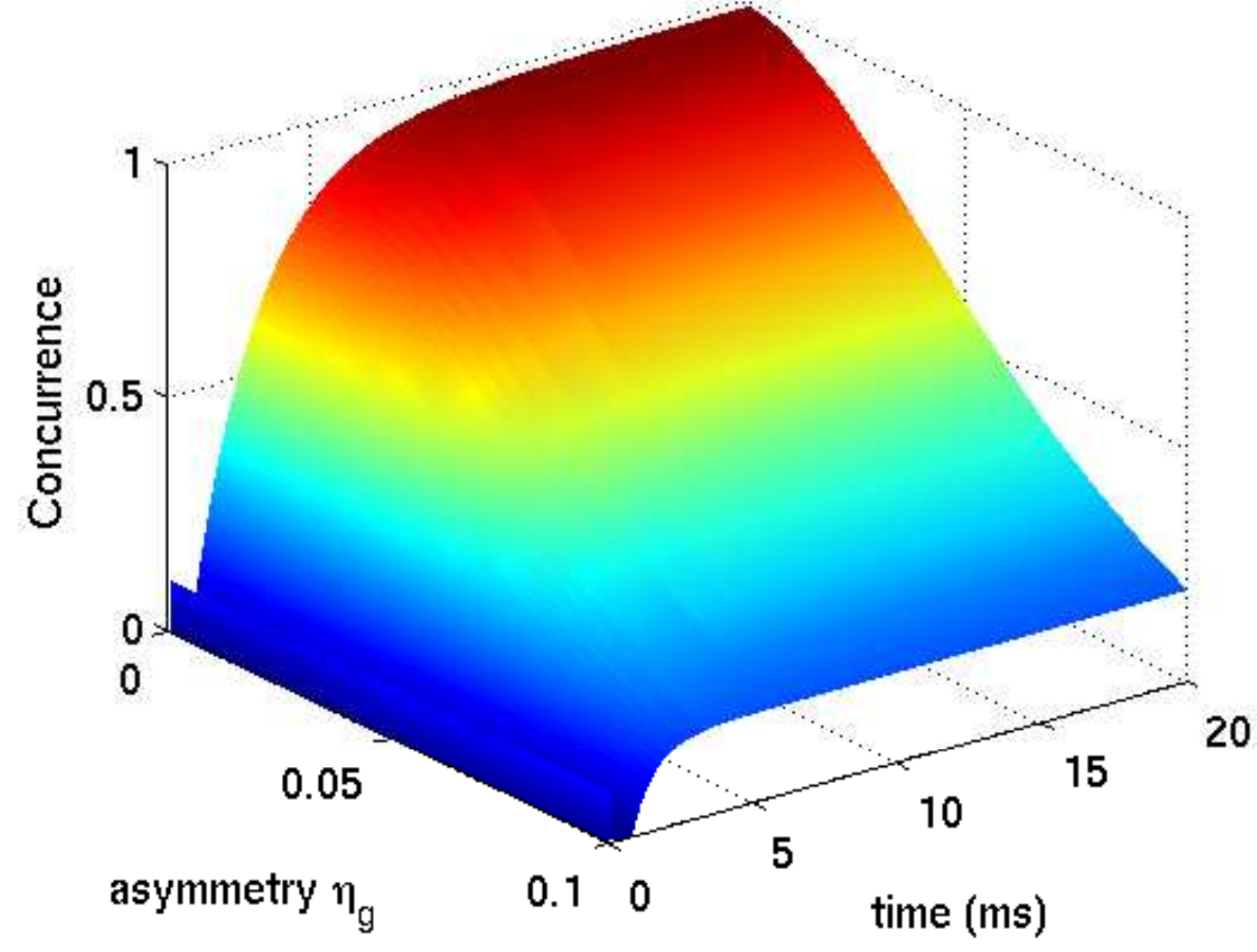}
\includegraphics[width=0.48\columnwidth]{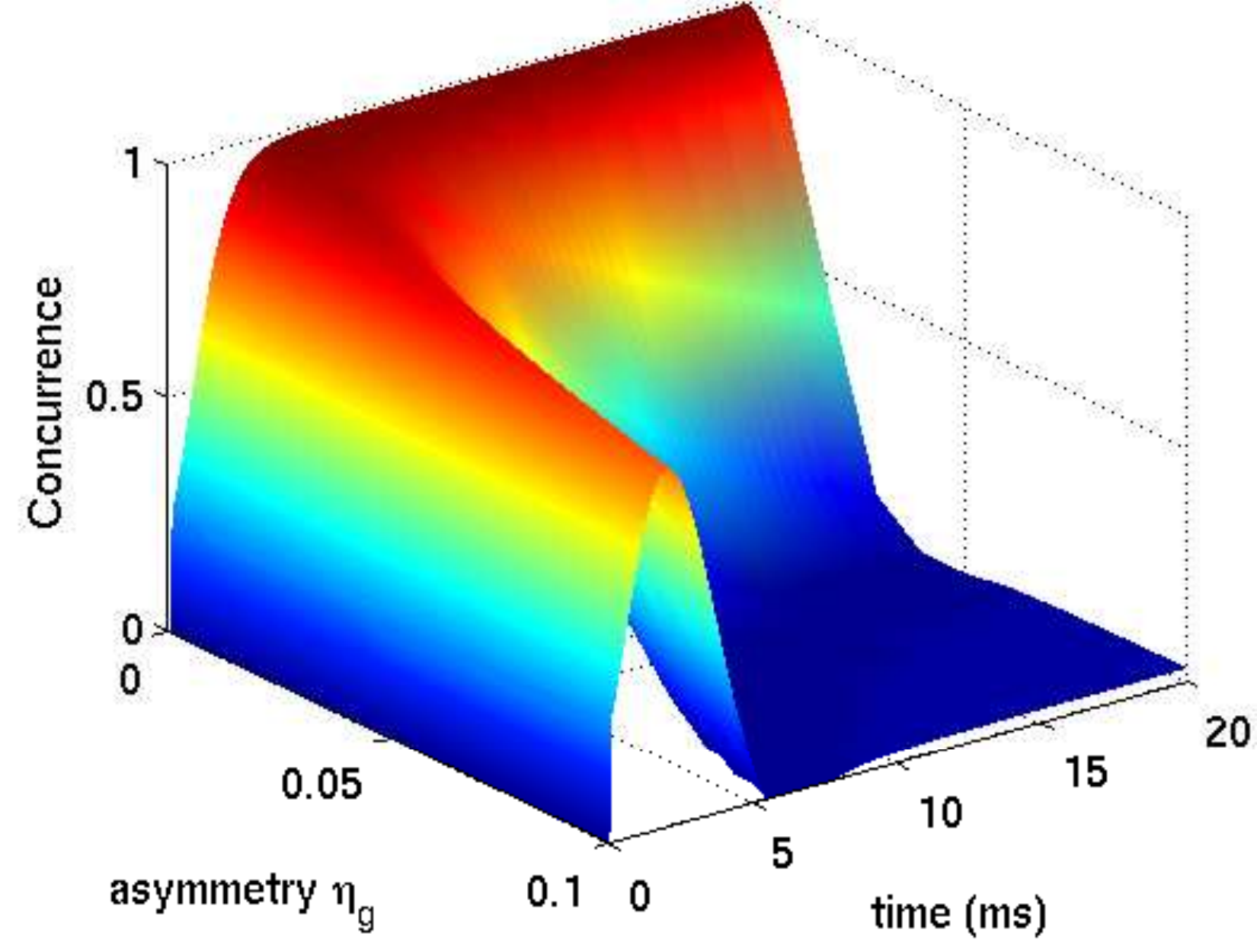}
\caption{(Color online): Concurrence $C(t)$ as a function of time
and asymmetry in the atom-cavity couplings $\eta_g$ for
$\Delta\omega=5.6$~kHz (left) and $\Delta\omega(t)=100 e^{-0.8t}$
(right).} \label{fig6}
\end{figure}

Another possible source of error is asymmetry in the couplings of the
atoms to the cavity.  If the atoms are identical and interact in phase
with the cavity field, i.e., if they are well localized in the cavity,
$g_1=g_2$ should be a good approximation even in the presence a small
magnetic field gradient $\vec B$ as $g$ is proportional to the magnetic
dipole moment, which is not affected by a small magnetic field gradient.
If there is an asymmetry, say $g_2=(1+\eta_g)g_1$, the system no longer
has a pure steady state but a unique mixed state in the interior, which
has lower concurrence than the MES, as illustrated in Fig.~\ref{fig6}.
Again, as for atomic decay, the exponential detuning profile allows us
to reach higher concurrences fast but the concurrence decreases as the
detuning vanishes.  Fig.~\ref{fig5} (right) shows that the results can
again be substantially improved by adding an offset to the exponential
detuning to stabilize the concurrence near its peak value.

Finally, we consider the effect of a frequency-offset that might result
if the atoms are not positioned symmetric with respect to the magnetic
field gradient, leading to asymmetric detuning and an error term
$H_e=\eta_\oo\Delta\oo(\s1^\dag\s1+\s2^\dag\s2)$.  Surprisingly high
entanglement is attainable even for large $\eta_\oo$.  A numerical fit
to the simulation data shows that the concurrence after $20$~ms is well
described by a quadradic model
$C_{20}(\eta_\oo)=a_2\eta_\oo^2+a_1\eta_\oo+a_0$ with
$(a_2,a_1,a_0)=(-0.062,-0.0024,0.99)$ for $\Delta\omega=5.6$~kHz and
$(-0.15,-0.039,1.00)$ for $\Delta\oo(t)=100e^{-0.8t}$.  For
$\eta_\oo>0.2$ the constant detuning is preferable as the large initial
amplitude of the exponential detuning magnifies the error, but in both
cases substantial entanglement, $92.66$\% and $81.09$\%, is still
attainable even if the offset is on the order of the detuning,
$\eta_\oo=1$.

We have shown that effectively maximal entanglement of non-interacting
atoms can be realized by putting the atoms in a lossy cavity driven by a
classical field and applying a magnetic field gradient.  While there are
many schemes for entangling qubits, this scheme exploits a novel concept
of \emph{environment-induced entanglement} to create \emph{high-grade
entanglement} and offers exceptional \emph{simplicity}, eliminating the
need for direct coupling between the atoms, control lasers to manipulate
internal atomic states, and complicated measurement setups and problems
due to limited detection efficiency.  Global attractivity of the target
state ensures that the atoms converge to this state for any initial
state, eliminating the need for initial state preparation.  Error
analyis shows that high-grade entanglement can be still achieved with
simple detuning profiles in the presence of atomic decay, asymmetric
atom-cavity coupings or frequency offsets.

We thank M.~Atat\"ure, C.~Lu, C.~Zipkes, C.~Sias from the Cavendish
Laboratory, and S.~Bose and L.-C.~Kwek for valuable discussions, and
acknowledge funding from EPSRC ARF Grant EP/D07192X/1 and Hitachi.

\end{document}